%% file: main.tex
\documentclass[submission, Phys]{SciPost}

\usepackage[utf8]{inputenc}
\usepackage{booktabs}
\usepackage[multiple]{footmisc}
\usepackage{lipsum}
\usepackage{rotating}
\usepackage{perpage}
\usepackage{chronology}
\usepackage{lipsum}
\usepackage{csquotes}

\usepackage{amssymb}
\usepackage{amsbsy}
\usepackage{amsmath}
\usepackage{abraces}
\usepackage{siunitx}
\usepackage{physics}
\usepackage[version=4]{mhchem}
\usepackage{bm}
\usepackage{bbm}
\usepackage{xfrac}
\usepackage{mathtools}

\usepackage{tikz}
\usepackage[T1]{fontenc}
\usepackage{etoolbox}
\usepackage{graphics}
\usepackage{multirow}
\usepackage{url}
\usepackage{hyperref}
\usepackage{placeins}
\usepackage[normalem]{ulem}

\usepackage{color}
\def\blue#1{{\color{blue}{#1}}}

\usepackage{cleveref}
\crefname{section}{Sec.}{Secs.}
\crefname{figure}{Fig.}{Figs.}
\crefname{table}{Tab.}{Tabs.}
\crefname{equation}{Eq.}{Eqs.}
\crefname{appendix}{App.}{Apps.}
\def\refcite#1{Ref.~\cite{#1}} 

\input{commands.tex}

\newcommand{\dsq}{\phantom{}^2}

\begin{document}

\begin{center}{\Large \textbf{
Simulating the interplay of dipolar and quadrupolar interactions in NMR by spin dynamic mean-field theory
}}\end{center}

\begin{center}
Timo Gräßer\textsuperscript{1,*},
Götz S. Uhrig\textsuperscript{1},
\end{center}

\begin{center}
{\bf 1} Condensed Matter Theory, TU Dortmund University, Otto-Hahn Stra\ss{}e 4, 44221 Dortmund, Germany
\\
* timo.graesser@tu-dortmund.de
\end{center}

\begin{center}
\today
\end{center}


\section*{Abstract}
{\bf
The simulation of nuclear magnetic resonance (NMR) experiments is a notoriously difficult task, if many spins participate in the dynamics. The recently established dynamic mean-field theory for high-temperature spin systems (spinDMFT) represents an efficient yet accurate method to deal with this scenario. SpinDMFT reduces a complex lattice system to a time-dependent single-site problem, which can be solved numerically with small computational effort. Since the approach retains local quantum degrees of freedom, a quadrupolar term can be exactly incorporated. This allows us to study the interplay of dipolar and quadrupolar interactions for any parameter range, i.e., without the need for a perturbative treatment. We obtain a remarkable agreement with experimental data for an aluminium nitride monocrystal, which strongly suggests the use of spinDMFT as a prediction tool. Furthermore, we draw a comparison between a quantum-mechanical and a classical version of spinDMFT showing that local quantum effects are of great importance for the studied type of system.
}

\vspace{10pt}
\noindent\rule{\textwidth}{1pt}
\tableofcontents\thispagestyle{fancy}
\noindent\rule{\textwidth}{1pt}
\vspace{10pt}

\section{Introduction}
\label{sec:intro}

Atomic nuclei with a spin $S>\sfrac12$ experience an electric quadrupolar interaction in an anisotropic electronic environment. This is relevant in the broad field of nuclear magnetic resonance (NMR) \cite{Slichter,ashbr09}, but also for the spin dynamics in charged quantum dots \cite{bulut12,bulut14,hackm15,glase16,joch23,Glazov}. Discovered in the 1930s, NMR has continuously evolved into a widely used technique in material physics, chemistry, biology and medicine. The principle of NMR is to resonantly address the nuclear spins of a sample in order to gain information about the sample's composition and molecular structure. This is based on the fact that nuclear spins interact with their chemical environment through the associated magnetic moments. Important contributions to the nuclear spin dynamics include the chemical shift, the J-coupling, the dipole-dipole interaction and the aforementioned quadrupolar interaction \cite{Levitt}. The latter two are the focus of this article. The dipolar interaction couples nearby nuclear spins to one another proportional to $1/r^3$, where $r$ is the relative distance. This implies a many-particle problem. The quadrupolar interaction, on the other hand, is completely local. It results from a coupling between the electric quadrupole moment of a deformed atomic nucleus with an electric field gradient generated by the surrounding electron cloud. 

In general, the simulation of a large spin system represents a notoriously difficult task. As the Hilbert space grows exponentially with the system size, exact simulations \cite{weiss06} are only feasible for a few tens of spins and therefore suffer from finite-size effects. In many systems, dipolar interactions are well captured by classical simulations \cite{elsay15,stark18}. In this case, the computational effort grows only polynomially with the system size so that, in practice, finite-size effects can be essentially removed. Despite this clear advantage, it is not \emph{a priori} clear how well the classical approximation works in a specific geometry. The accuracy is expected to be reduced in low-dimensional systems or systems with well-separated, small groups of spins, where quantum effects tend to be more relevant. Hybrid quantum-classical approaches can assist to some degree \cite{stark20}, but as they do not make use of translational invariance, they can become quite demanding. Besides this, it is not clear, how well an additional quadrupolar interaction can be treated in classical or hybrid simulations. The key question is how important the local quantum nature of the spins is.

In many scenarios, the quadrupolar interaction strongly dominates the dipolar one \cite{Levitt}. If, in addition, the quadrupolar coupling varies from nucleus to nucleus in the sample, the dipolar contribution to the line shapes is not visible and can essentially be neglected. But this is not always the case. Prominent counter-examples include samples containing \ce{^{7}Li} \cite{eliav96,eliav13}, where homonuclear dipolar interactions have been observed to significantly affect stimulated-echo spectra \cite{qi02,qi04}. Often, a perturbative treatment of the dipolar interaction \cite{dieze95a,dieze95b} or an exact simulation of a few adjacent spins \cite{wi00,qi04} suffices to capture the main physics. However, it is not clear how reliable such approaches are if the quadrupolar and dipolar interaction are of the same order of magnitude.

In this article, we introduce spin dynamic mean-field theory, short spinDMFT, as an alternative approach for simulating the interplay of dipolar and quadrupolar interactions. SpinDMFT is developed for dense spin systems at infinite temperature \cite{graes21}. \enquote{Dense} in this context means that the approach is accurate in the limit where each spin has an infinite number of interaction partners. \enquote{Infinite temperature} corresponds to the thermal energy being much larger than any internal energy scale of the considered system. Then, the initial statistical operator corresponds to completely disordered spins. On the one hand, this is a strong constraint, but on the other hand it makes spinDMFT perfectly tailored to the field of NMR because nuclear spins are disordered in most experiments due to the smallness of their gyromagnetic ratios. A strong advantage of spinDMFT is that it requires only small computational effort, which allows for systematic extensions such as cluster spinDMFT \cite{graes23} and non-local spinDMFT \cite{graes24}. Moreover, the method is highly versatile because it works with an effective single-site Hamiltonian. This easily allows for the inclusion of local spin terms such as local magnetic fields, static or time dependent, as well as quadrupolar interactions. 

The article is set up as follows. First, we formulate a basic model for a spin system containing a dipolar and quadrupolar interaction in \cref{sec:basicmodel}. Subsequently, in \cref{subsec:spinDMFT:theory}, we apply spinDMFT to this model obtaining a single-site model that can be solved numerically. The results are presented and discussed in Secs.~\ref{subsec:spinDMFT:results_time} and \ref{subsec:spinDMFT:results_freq}. In \cref{sec:experiment}, we benchmark the method by a comparison with experimental results for an aluminium nitride (AlN) monocrystal. In \cref{sec:classspinDMFT}, we draw a comparison to a classical analogue system. Finally, the article is concluded in \cref{sec:conclusion}.

\section{Basic model}
\label{sec:basicmodel}

We consider a high-temperature nuclear spin ensemble of homogeneous, spatially-fixed spins with $S>\sfrac12$. The ensemble shall be subject to a strong magnetic field as usual in NMR experiments. The spins interact with one another via the secular homonuclear Hamiltonian \cite{Levitt}
\begin{align}
    \oH_{\mathrm{DD}} &= \frac12 \sum_{i,j} d_{ij} \left( 2 \Szi \Szj - \Sxi \Sxj - \Syi \Syj \right).
    \label{eqn:dipoledipole}
\end{align}
with
\begin{align}
    d_{ij} &\coloneqq d_{\vec{r}_{ij}}(\nB) = \frac{1 - 3 \left(\vec{n}_{ij}\cdot\nB\right)^2}{2} \frac{\mu_0}{4\pi} \frac{\gamma_i \gamma_j \hbar}{|\vec{r}_{ij}|^3}, &
    \vec{n}_{ij} &\coloneqq \frac{\vec{r}_{ij}}{|\vec{r}_{ij}|}, &
    \nB &\coloneqq \frac{\vec{B}}{|\vec{B}|},
\end{align}
where $\vec{r}_{ij}=\vec{r}_{j}-\vec{r}_{i}$ is the distance vector between spins $i$ and $j$ and $\vec{B}$ is the magnetic field. Any self-interactions are ruled out, i.e., we set $d_{ii}\coloneqq 0$. In \cref{eqn:dipoledipole} and henceforth, we label operators by boldface symbols. 
Each nucleus locally interacts with an electric field gradient (EFG), which is captured by the secular quadrupolar interaction term \cite{Levitt}
\begin{align}
    \oH_{\mathrm{Q}} &= \Omega \sum_i \left( 3 \Szi\dsq - \vSi^2 \right)
\end{align}
with
\begin{align}
   \Omega &\coloneqq \frac{\chi}{2S(2S-1)} V^{zz}(\nB),
\end{align}
where $\chi$ is the electric quadrupole moment of the nucleus and $V^{zz}$ is the $zz$-component of the EFG tensor. The square of the spin vector operator is a constant and thus corresponds to a constant energy shift, which is irrelevant for the spin dynamics and will be omitted henceforth. The Hamiltonian 
\begin{align}
    \oH &= \oH_{\mathrm{DD}} + \oH_{\mathrm{Q}}
\end{align}
describes a complex many-body quantum system, which cannot be exactly solved for large numbers of spins. An approximation has to be made.

\section{Spin dynamic mean-field theory}

\subsection{Single-site model and closed self consistency}
\label{subsec:spinDMFT:theory}

Spin dynamic mean-field theory is an elegant and efficient way to describe this many-body system approximately \cite{graes21}. The approximation is reliable if each spin has a large number of interaction partners. This implies that the fields describing the local spin environments (henceforth called local-environment fields) defined by
\begin{align}
    \vV_i(t) &\coloneqq \sum_{j} d_{ij} \dul{D} \, \vSj, & \dul{D} &= 
    \begin{pmatrix}
        -1 & 0  & 0 \\
        0  & -1 & 0 \\
        0  & 0  & 2 \\
    \end{pmatrix},
\end{align}
consist of many contributions. A measure for the number of contributions is the effective coordination number \cite{graes21}
\begin{align}
    \zeff \coloneqq \frac{\left( \sum_{j}d_{ij}^2 \right)^2}{ \sum_{j}d_{ij}^4 },
\end{align}
which is independent of $i$ as the system is homogeneous. If $\zeff$ is large ($\gtrapprox 5$), it is well-justified to replace each local-environment field by a dynamic Gaussian mean-field $\cvV(t)$. This results in a local mean-field Hamiltonian
\begin{align}
    \oH^{\mf}(t) &= \cvV(t) \cdot \vS + 3 \Omega \Sz\dsq.
    \label{eqn:mfHam}
\end{align}
The mean-field is zero on average and its second moments result from the self-consistency condition \cite{graes21}
\begin{align}
    \mfav{ \cVa(t) \cVb(0) } &= \JQ^2 \delta^{\alpha\beta} D^{\alpha\alpha}\dsq \expval{ \Sa(t) \Sa(0) },
    \label{eqn:selfcons}
\end{align}
where $D^{\alpha\alpha}$ are diagonal matrix elements of $\dul{D}$ and we defined the quadratic coupling constant
\begin{align}
    \JQ^2 &\coloneqq \sum_{j} d_{ij}^2.
    \label{eqn:quadcoup}
\end{align}
The formal equation to compute the spin autocorrelations is given by 
\begin{align}
    \expval{ \Sa(t) \Sa(0) } &= \int \mathrm{D}\mathcal{V} \, p(\mathcal{V}) \,\expval{ \Sa(t) \Sa(0) }^{\text{loc}}(\mathcal{V}).
    \label{eqn:mfspinautocorr}
\end{align}
Here, $\mathcal{V}$ is a mean-field time series and $p(\mathcal{V})$ its multivariate Gaussian probability distribution. The expectation value $\expval{ \Sa(t) \Sa(0) }^{\text{loc}}(\mathcal{V})$ is carried out in the Hilbert space of a single spin considering the time evolution generated by the Hamiltonian in \cref{eqn:mfHam} for a specific $\mathcal{V}$. Since the system is at infinite temperature, the density matrix is proportional to the identity in all considered expectation values. 

The defined self-consistency problem is solved by numerical iteration. Starting from an initial guess for the spin autocorrelations, one computes the second mean-field moments via \cref{eqn:selfcons} and uses them to update the spin autocorrelations by means of \cref{eqn:mfspinautocorr}. This process is repeated until the spin autocorrelations are converged, which requires only about 5 iteration steps. In practice, the path integral in \cref{eqn:mfspinautocorr} is evaluated by discretizing the time and applying a Monte-Carlo simulation. For more details on the derivation and numerical implementation of the approach, we refer to the original article in \refcite{graes21}.

\subsection{Results in the time domain}
\label{subsec:spinDMFT:results_time}

\begin{figure}[h]
    \centering
    \includegraphics[width=\textwidth]{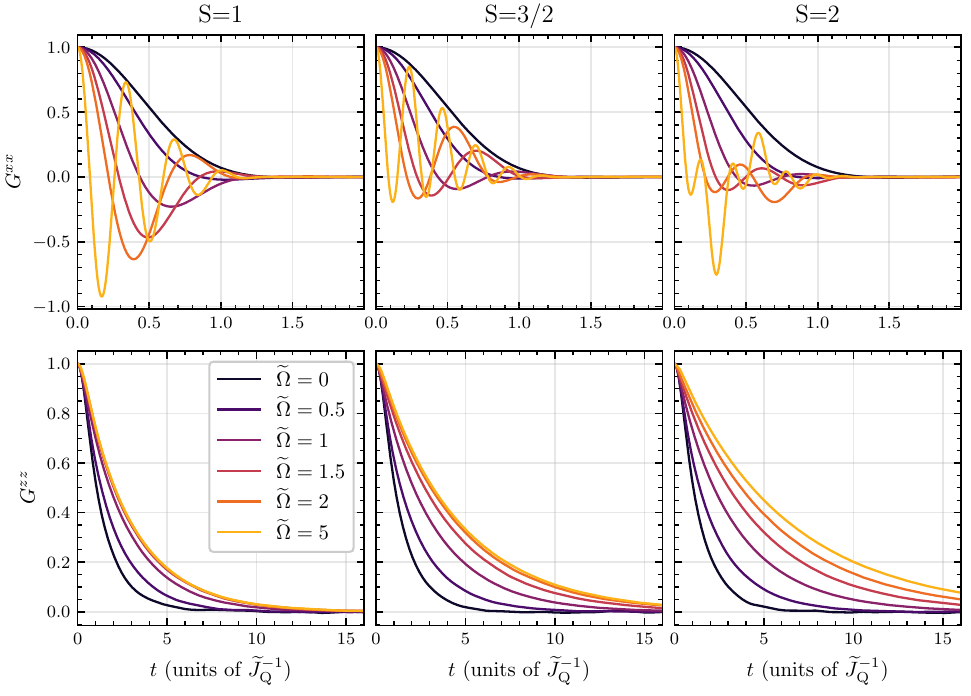}
    \caption{Results for the normalized spin autocorrelation $G^{\alpha\alpha}$ from spinDMFT for different spin lengths and quadrupolar interaction strengths in the time domain. The top row shows the transverse and the bottom row the longitudinal results. The spin length is increased from left to right. Different quadrupolar strengths are indicated by different colors according to the provided legend. Numerical errors are of the order $\SI{1}{\percent}$ or smaller of the signal amplitude at $t=0$.}
    \label{fig:spinDMFTtimeresults}
\end{figure}

Figure \ref{fig:spinDMFTtimeresults} displays the converged results of the normalized autocorrelations
\begin{align}
    G^{\alpha\alpha}(t) &\coloneqq \frac{3}{S(S+1)} g^{\alpha\alpha}(t),
\end{align}
with
\begin{align}
    g^{\alpha\alpha}(t) &\coloneqq \expval{ \Sa(t) \Sa(0) }
\end{align}
for different spin lengths $S$ and quadrupolar interaction strengths in the time domain. 

The top panels each show the transverse autocorrelation $G^{xx}=G^{yy}$ and the bottom panels the longitudinal autocorrelation $G^{zz}$. Any off-diagonal autocorrelations vanish due to rotational symmetry about the $z$-axis. The time is given in units of $1/\widetilde{J}_{\text{Q}}$ defining
\begin{align}
    \widetilde{J}_{\text{Q}} \coloneqq \sqrt{\frac{S(S+1)}{3}} \, \JQ,
\end{align}
where $\JQ$ is the quadratic coupling constant defined in \cref{eqn:quadcoup} and $\hbar$ is set to one. We choose this specific spin-dependent rescaling for better comparison of the results of different spin lengths. Increasing $S$ means increasing the strength of the mean-field and thus the speed of the decay. This scaling effect is compensated when depicting the time in units of $1/\widetilde{J}_{\text{Q}}$. Yet, the results for different spin lengths are visibly different. The quadrupolar interaction leads to oscillations in the transverse autocorrelations. This behavior is similar to a Larmor precession due to a magnetic field in $z$-direction. However, it is clearly not the same as the quadrupolar interaction term is quadratic in $\Sz$ in contrast to the Zeeman term. The oscillation frequencies depend on the quadrupolar interaction strength 
\begin{align}
    \widetilde{\Omega} \coloneqq \frac{\Omega}{\JQ}.
\end{align}
In the case of $S=2$, two oscillations with different frequencies are overlapping. This behavior is best understood in the spectra, which will be considered in the next subsection. The longitudinal results do not oscillate, but nevertheless depend on the quadrupolar interaction strength. They show a monotonic decay which slows down upon increasing $\widetilde{\Omega}$. This is not surprising as the quadrupolar interaction destabilizes the transverse spin components so that they average out faster. In return, the longitudinal correlation decays slower because its decay is driven by the transversal components. A qualitatively similar behavior was obtained when adding a static Gaussian noise in the $z$-direction, see \refcite{graes21}.

\begin{figure}[h]
    \centering
    \includegraphics[width=\textwidth]{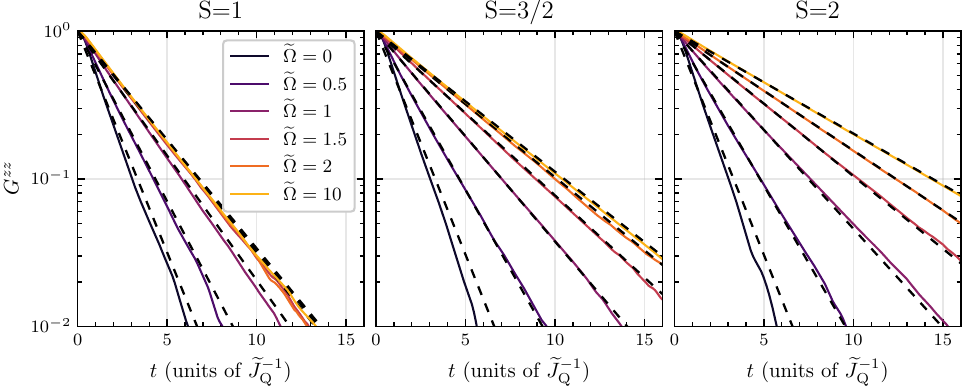}
    \caption{Results of the longitudinal spin autocorrelations in the time domain in logarithmic representation. The dashed lines display exponential fits, which work exceptionally well. Numerical errors are of the order $\SI{1}{\percent}$ or smaller of the signal amplitude at $t=0$. The wiggles of the data, which become visible for small $G^{zz}$, result from the statistical error of the Monte-Carlo simulation.}
    \label{fig:longexpfit}
\end{figure}

For large values of $\widetilde{\Omega}$, the longitudinal correlations depend only weakly on $\widetilde{\Omega}$. We analyze this further by fitting exponentials $\mathrm{exp}(-t/T)$ to the results. The fits work moderately well for small $\widetilde{\Omega}$ and exceptionally well for large $\widetilde{\Omega}$, see \cref{fig:longexpfit} and \cref{tab:longexpfit}.

\begin{table}[h]
    \centering
    \caption{Extracted decay times of the longitudinal autocorrelations $G^{zz}$ in dependence of $\widetilde{\Omega}$ for different spin lengths. The corresponding exponential fits are shown in \cref{fig:longexpfit}. The provided errors result from the finite time discretization, which is the dominant source of numerical error. For small $\widetilde{\Omega}$ the error of the fit procedure is larger than the discretization error.}
    \label{tab:longexpfit}
    \begin{tabular}{c|c|c|c}
     & $S=1$ & $S=3/2$ & $S=2$ \\
    $\widetilde{\Omega}$ & $T / \widetilde{J}_{\mathrm{Q}}^{-1}$& $T / \widetilde{J}_{\mathrm{Q}}^{-1}$& $T / \widetilde{J}_{\mathrm{Q}}^{-1}$\\
    \hline
    0 & 1.45(1) & 1.434(1) & 1.432(3) \\
    0.5 & 1.883(6) & 2.020(6) & 2.073(2) \\
    1 & 2.57(1) & 3.05(1) & 3.24(1) \\
    1.5 & 2.88(1) & 3.90(7) & 4.42(6) \\
    2 & 2.92(5) & 4.39(7) & 5.4(1) \\
    2.5 & 2.93(4) & 4.50(5) & 5.89(2) \\
    3 & 2.94(2) & 4.54(5) & 6.13(5) \\
    5 & 2.95(5) & 4.6(1) & 6.3(2) \\
    10 & 2.95(5) & 4.6(1) & 6.3(2) \\
    \end{tabular}
\end{table}

\begin{figure}[h]
    \centering
    \includegraphics[width=0.6\textwidth]{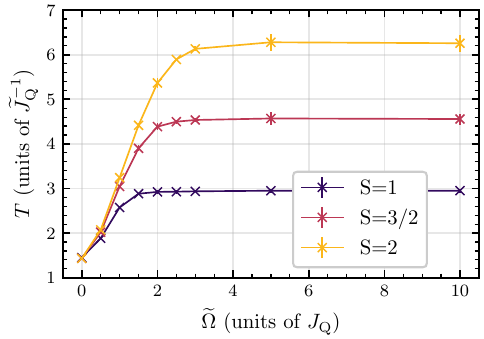}
    \caption{Plot of the extracted decay times of the longitudinal autocorrelations $G^{zz}$ in dependence of $\widetilde{\Omega}$ for different spin lengths. The corresponding exponential fits are shown in \cref{fig:longexpfit}. The error bars result from the finite time discretization, which is the dominant source of numerical error.}
    \label{fig:long_decay_times}
\end{figure}

The extracted relaxation times reach a spin-length dependent saturation value for \mbox{$\widetilde{\Omega}~\to~\infty$} as can be seen in \cref{fig:long_decay_times}. This is evidence that the system can be described by an effective Hamiltonian in this limit. Specifically for spin $S=1$, we obtain
\begin{subequations}
\begin{align}
    \oH_{\eff,ij} &= \overline{\oUd_{\mathrm{Q},ij}(t) \oH_{ij} \oU_{\mathrm{Q},ij}(t) - \oH_{\mathrm{Q},ij}} \\
    &= d_{ij} \left[ 2 \Szi \Szj - \frac12 \left(\Sxi \Sxj + \Syi \Syj + \{\Sxi,\Szi\}\{\Sxj,\Szj\} +\{\Syi,\Szi\}\{\Syj,\Szj\} \right)\right],
\end{align}
\end{subequations}
by average Hamiltonian theory \cite{blane09} similar to the secular approximation \cite{Levitt}. The overline denotes a time average, the curly brackets denote the anticommutator and we defined
\begin{subequations}
\begin{align}
    \oH_{\mathrm{Q},ij} &\coloneqq 3 \Omega \left( \Szi\dsq + \Szj\dsq \right), \\
    \oU_{\mathrm{Q},ij}(t) &\coloneqq \exp{ \im \oH_{\mathrm{Q},ij} t}, \\
    \oH_{ij} &\coloneqq d_{ij} \left( 2 \Szi \Szj - \Sxi \Sxj - \Syi \Syj \right) + \oH_{\mathrm{Q},ij}.
\end{align}
\end{subequations}
It is important to note that this effective Hamiltonian results specifically for $S=1$ and has a different form for different spin lengths. The effective Hamiltonian may be treated directly by spinDMFT, but this is beyond the scope of the present article.

\subsection{Results in the frequency domain}
\label{subsec:spinDMFT:results_freq}

\begin{figure}[h!]
    \centering
    \includegraphics[width=\textwidth]{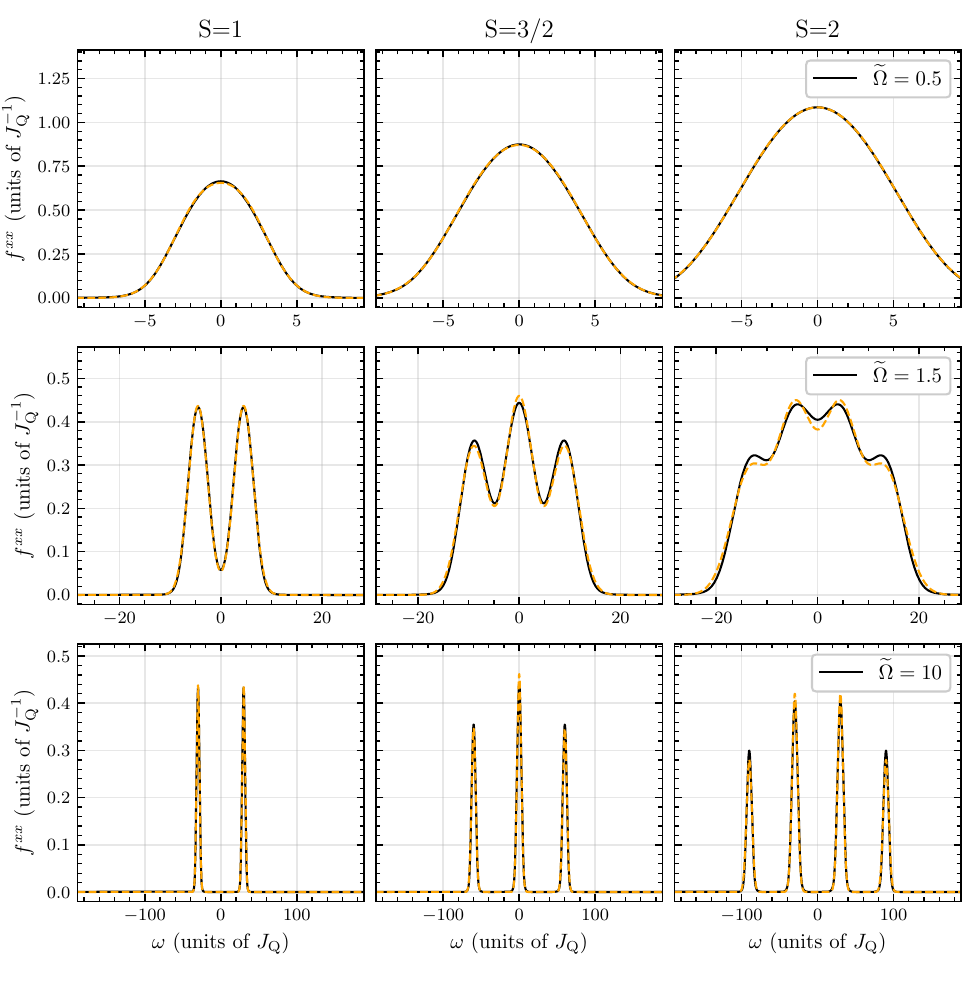}
    \caption{Fourier transform $f^{xx}(\omega)\coloneqq \int_{-\infty}^{\infty} \mathrm{e}^{-\im \omega t} g^{xx}(t) \mathrm{d}t$ of the transverse spin autocorrelation $g^{xx}(t)$ for different quadrupolar interaction strengths and spin lengths. The spin length is increased from left to right and the quadrupolar interaction from top to bottom. The orange dashed line corresponds to the Gaussian fit described in \cref{eqn:fitfunction}. Small deviations are seen at some of the peak maxima. These become smaller when allowing for an individual amplitude $A_i$ for each peak in the fit function. However, we prefer the shown fits because they require only a single parameter, namely, the standard deviation $\sigma$. The numerical errors of the simulation data are smaller than the width of the lines.}
    \label{fig:spectra}
\end{figure}

The spectra are shown in \cref{fig:spectra} for different spin lengths and strengths of the quadrupolar coupling. They are obtained by fast Fourier transform of the symmetrized temporal results. 

We stress that the spectra are not referring to the free-induction decay (FID), but to the spin autocorrelation. The latter can be considered a first-order approximation of the FID, which corresponds to a superposition of the autocorrelation with pair correlations. These are not directly accessible in single-site spinDMFT, but may be computed by the extension non-local spinDMFT \cite{graes24}. This is beyond the scope of the present article. However, it is worth mentioning that the inclusion of a quadrupolar coupling makes the spin dynamics more local so that paircorrelations become less relevant with respect to the autocorrelation. We therefore consider the autocorrelation spectra to be a good approximation of the FID for large $\Omega$.

\begin{figure}[h]
    \centering
    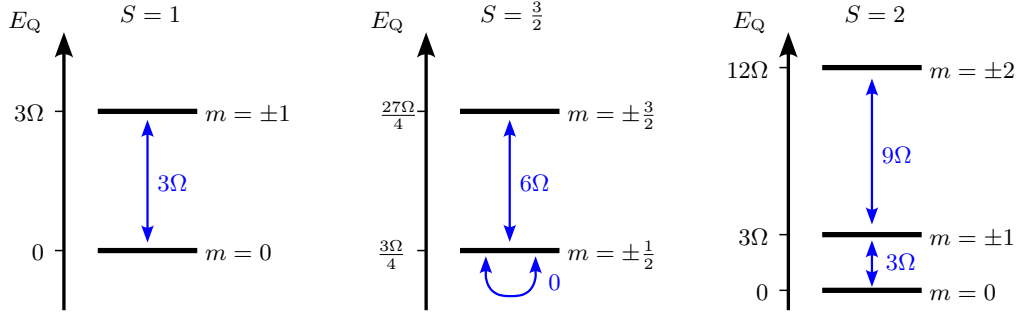
    \caption{Possible quadrupolar transitions with $\Delta m = \pm 1$ for different spin lengths. The quadrupolar energy is given by $E_{\text{Q}}=3\Omega m^2$.}
    \label{fig:transitions}
\end{figure}

The peaks in \cref{fig:spectra} are located rather precisely at
\begin{align}
    \omega_{i} &= 6\Omega\left(S-i+\frac12\right), &i &\in \{1,\dots,2S\}.
\end{align}
This can be understood from the local quantum mechanics: The peaks correspond to the expected quadrupolar transitions, see \cref{fig:transitions} for visualization. The advantage of spinDMFT consists in the \emph{ab initio} prediction of the continuous line shapes induced by the dipolar interactions.

Remarkably, we find that spinDMFT predicts a Gaussian broadening of the resonance lines over the full considered parameter range. This can be well seen from the dashed lines in \cref{fig:spectra} which correspond to the fit function
\begin{align}
    \!\!\!f_{\sigma}(\omega) = \pi \sum_{i=1}^{2S} \frac{i}{2} \left(1-\frac{i}{2S+1} \right) \frac{ \text{exp}\left( - \frac{(\omega-3\Omega(2S-2i+1))^2}{2\sigma^2} \right) + \text{exp}\left( - \frac{(\omega+3\Omega(2S-2i+1))^2}{2\sigma^2} \right) }{\sqrt{2\pi\sigma^2}},
    \label{eqn:fitfunction}
\end{align}
which is derived from considering the exact local result and replacing the $\delta$-functions by Gaussian distributions. The fit parameters are listed in \cref{tab:spectra_fit_params}. To account for the main effect of the spin length, the standard deviations are provided in units of $\widetilde{J}_{\text{Q}}$. In this unit, they depend only weakly on the spin length $S$ and the quadrupolar interaction strength $\widetilde{\Omega}$. A Gaussian line broadening results for static Gaussian-distributed Ising mean-fields \cite{Graesser_PhD} and zero quadrupolar interactions\footnote{By Gaussian-distributed mean-fields, we refer to the distribution functional from which the mean-fields are drawn and not to their correlation in time. The latter is constant in case of static mean-fields.}. SpinDMFT predicted that this behavior is stable upon introducing transverse interactions as they occur in a secular dipolar Hamiltonian \cite{graes21}. As we obtain here, an additional quadrupolar interaction does not affect the line broadening qualitatively. 


The longitudinal autocorrelations are well captured by exponential fits in the time domain, see \cref{fig:longexpfit}. Therefore, their spectra are very well described by Lorenz curves $\Gamma/(\Gamma^2 +\omega^2)$ with decay rates $\Gamma = 1/T$. We refrain from showing the corresponding plots. The decay times $T$ are shown in \cref{tab:longexpfit}. The longitudinal autocorrelations are important for the spin dynamics, but difficult to access experimentally, because one would need to polarize an individual spin in a homogeneous ensemble. In principle, the same holds for the transverse autocorrelation, but as we mentioned before, the latter is not much different from the experimentally accessible FID. Hence, we consider a comparison to experimentally measured FID's in the following section.

\begin{table}[h]
    \centering
    \caption{Resulting fit parameters for the spectra shown in \cref{fig:spectra} corresponding to the fit function in \cref{eqn:fitfunction}. The provided errors result from the finite time discretization, which is the main source of numerical error.}
    \begin{tabular}{c|c|c|c}
        & $S=1$ & $S=3/2$ & $S=2$ \\
        $\widetilde{\Omega}$ & $\sigma / \widetilde{J}_{\mathrm{Q}}$& $\sigma / \widetilde{J}_{\mathrm{Q}}$& $\sigma / \widetilde{J}_{\mathrm{Q}}$\\
        \hline
        0.5 & 2.21(1) & 2.210(3) & 2.208(2) \\
        1.5 & 2.343(8) & 2.456(5) & 2.53(1) \\
        10 & 2.335(3) & 2.43(1) & 2.530(9) \\
    \end{tabular}
    \label{tab:spectra_fit_params}
\end{table}

\section{Comparison to experimental data}
\label{sec:experiment}

\subsection{Test system}

\begin{figure}[h]
    \centering
    \includegraphics[width=0.6\textwidth]{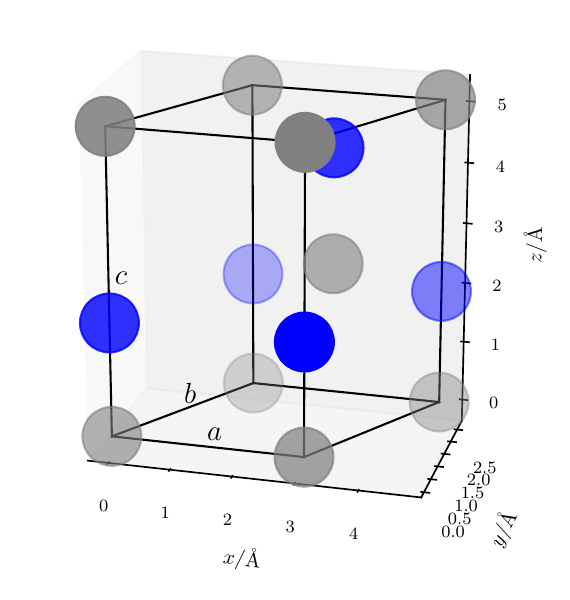}
    \caption{Schematic representation of the unit cell of the wurtzite crystal structure with two different atomic species, here aluminium (grey) and nitrogen (blue). The labels a, b and c denote the crystallographic axes.}
    \label{fig:AlN_structure}
\end{figure}

We compare our approach with experimental data from \refcite{zeman20} measuring FID signals of an aluminium nitride (AlN) monocrystal. AlN is a salt crystal composed of aluminum and nitrogen ions arranged in a hexagonal wurtzite structure. The unit cell is shown schematically in \cref{fig:AlN_structure}. 

The isotopes \ce{^{27}Al} and \ce{^{14}N} make up for about $\SI{100}{\percent}$ and $\SI{99.6}{\percent}$, respectively, of the natural abundance. Both are NMR active carrying a nuclear spin of $S=5/2$ (\ce{^{27}Al}) and $I=1$ (\ce{^{14}N}) with gyromagnetic ratios of $\gamma_{\Al} = \SI{11.103}{\mega\hertz\per\tesla}$ and $\gamma_{\Ni} = \SI{3.077}{\mega\hertz\per\tesla}$. With a length of $r\approx\SI{1.90}{\angstrom}$, the Al-N bonds along the crystallographic c axis are slightly larger than the remaining bonds with $r\approx\SI{1.89}{\angstrom}$ \cite{zeman20}. This induces an EFG at the nuclear sites, which leads to strong quadrupolar interactions. The quadrupolar coupling constants have been measured in \refcite{zeman20} to be $\chi_{\Al} \approx \SI{1.914}{\mega\hertz}$ and $\chi_{\Ni} \approx \SI{8.19}{\kilo\hertz}$. Since the principle-axis system and the crystal lattice frame coincide, the orientation-dependent EFG tensor can be deduced by
\begin{align}
    V^{zz}(\nB) &= \frac12 \left( 3 \cos^2 (\vartheta) - 1 \right),
    \label{eqn:efg}
\end{align}
where $\vartheta$ is the angle between the magnetic field direction and the crystallographic c axis. For the sake of completeness, we mention that the nuclear spins are also subject to chemical shifts. However, as they do not vary among spins of the same species, they only shift the spectra and do not affect the lineshapes. Hence, we neglect them in the modelling for simplicity.

\subsection{Adaptation of spinDMFT}

The AlN crystal structure contains two distinct aluminium and two distinct nitrogen sites per unit cell. While each two sites of the same spin species are equivalent with respect to quadrupolar interactions and chemical shifts, the non-local dipolar couplings may vary. An adequate treatment by spinDMFT requires thus to distinguish four different sites, which we denote $\Al\X$ and $\Ni\X$ with $\X \in \{A,B\}$. Analogous to \cref{eqn:mfHam}, the mean-field Hamiltonians are given by
\begin{align}
    \oH^{\mf}_{\Al\X}(t) &= \cvV_{\Al\X}(t) \cdot \vS_\X + 3 \Omega_{\Al} \Sz_\X\dsq, & \oH^{\mf}_{\Ni\X}(t) &= \cvV_{\Ni\X}(t) \cdot \vI_\X + 3 \Omega_{\Ni} \Iz_\X\dsq.
\end{align}
The mean-fields $\cvV_{\Al\X}$ and $\cvV_{\Ni\X}$ contain contributions from all four sites. However, the secular approximation reduces the interaction between different spin species to Ising couplings only. The adapted self-consistency conditions for this scenario read
\begin{subequations}
\label{eqn:selfcons_AlN}
\begin{align}
    \mfav{ \cVa_{\Al\X}(t) \cVa_{\Al\X}(0) } &= \sum_{\X'} \left(\JQ^{\Al\X,\Al\X'}\right)^2 \expval{ \Sa_{\X'}(t) \Sa_{\X'}(0) }, \\
    \mfav{ \cVz_{\Al\X}(t) \cVz_{\Al\X}(0) } &= \sum_{\X'} \left(\JQ^{\Al\X,\Al\X'}\right)^2 \expval{ \Sz_{\X'}(t) \Sz_{\X'}(0) } + \left(\JQ^{\Al\X,\Ni\X'}\right)^2 \expval{ \Iz_{\X'}(t) \Iz_{\X'}(0) }, \\
    \mfav{ \cVa_{\Ni\X}(t) \cVa_{\Ni\X}(0) } &= \sum_{\X'} \left(\JQ^{\Ni\X,\Ni\X'}\right)^2 \expval{ \Ia_{\X'}(t) \Ia_{\X'}(0) }, \\
    \mfav{ \cVz_{\Ni\X}(t) \cVz_{\Ni\X}(0) } &= \sum_{\X'} \left(\JQ^{\Ni\X,\Ni\X'}\right)^2 \expval{ \Iz_{\X'}(t) \Iz_{\X'}(0) } + \left(\JQ^{\Ni\X,\Al\X'}\right)^2 \expval{ \Sz_{\X'}(t) \Sz_{\X'}(0) },
\end{align}
\end{subequations}
with $\alpha\in\{x,y\}$. The defined quadratic coupling constants sum only over bonds connecting the involved sites, e.g., 
\begin{align}
    \left(\JQ^{\Al \mathrm{A},\Ni \mathrm{B}}\right)^2 &= \sum_{j} \left(d^{\Al \mathrm{A},\Ni \mathrm{B}}_{ij}\right)^2,
\end{align}
where $d^{\Al A,\Ni B}_{ij}$ is the dipolar coupling between the aluminium spin on site A of unit cell $i$ and the nitrogen spin on site B of unit cell $j$. Since each mean-field depends on the autocorrelations of all four sites, one has to solve all self-consistency conditions simultaneously. This entails simulating the four single-site problems at once, superposing the resulting autocorrelations according to \cref{eqn:selfcons_AlN} and repeating the procedure with updated mean-fields until convergence is reached. The resulting converged autocorrelations are averaged with respect to the two distinct sites per spin species to obtain the final results for aluminium and nitrogen.

The experiment was carried out for various orientations of the magnetic field with respect to the crystal lattice frame. For the quadrupolar coupling, only the angle $\vartheta$ matters as can be seen in \cref{eqn:efg}\footnote{We highlight the different naming of angles here and in \refcite{zeman20}. The crystal rotation angle is denoted $\vartheta$ in the present study and $\varphi$ in the experimental article.}. The dipolar couplings, however, are affected also by the azimuthal angle $\varphi$. The rotation axis in \refcite{zeman20} is perpendicular to c and to one of the axes a or b. Corresponding to \cref{fig:AlN_structure}, we consider the rotation axis to be parallel to y, but the magnitude of deviations from this axis is not entirely clear. Two exemplary quadratic coupling matrices for specific pairs of angles are provided in \cref{tab:quad_coup_matrix}.

\begin{table}[h]
  \centering
  \caption{Quadratic coupling matrices $\left(\JQ^{\mu\X,\nu\X'}\right)^2$ for two different pairs of angles, $\vartheta=\SI{45}{\degree}-\SI{0.65}{\degree},\,\varphi=\SI{0}{\degree}$ (left) and $\vartheta=\SI{45}{\degree}-\SI{0.65}{\degree},\,\varphi=\SI{30}{\degree}$ (right). All values are in units of $(2\pi\si{\kilo\hertz})^2$. The blue entries depend on the azimuthal angle $\varphi$.}
  \label{tab:quad_coup_matrix}
  \begin{minipage}{0.48\textwidth}
    \centering
    \begin{tabular}{ccccc}
      & Al,A & Al,B & N,A & N,B \\
      \toprule
      Al,A & 0.066 &  0.20 &   0.013 &    \blue{0.068} \\
      Al,B &  0.20 & 0.066 &    \blue{0.068} &   0.013 \\
      N,A & 0.013 & \blue{0.068} & 0.00030 & 0.00090 \\
      N,B & \blue{0.068} & 0.013 & 0.00090 & 0.00030 \\
      \bottomrule
    \end{tabular}
  \end{minipage}
  \hfill
  \begin{minipage}{0.48\textwidth}
    \centering
    \begin{tabular}{cccc}
      Al,A & Al,B & N,A & N,B \\
      \toprule
      0.066 &   0.20 &   0.013 &    \blue{0.13} \\
       0.20 &  0.066 &  \blue{0.0082} &   0.013 \\
      0.013 & \blue{0.0082} & 0.00030 & 0.00090 \\
       \blue{0.13} &  0.013 & 0.00090 & 0.00030 \\
      \bottomrule
    \end{tabular}
  \end{minipage}
\end{table}

\subsection{Results}

\begin{figure}[h!]
  \centering
  \includegraphics[width=\textwidth]{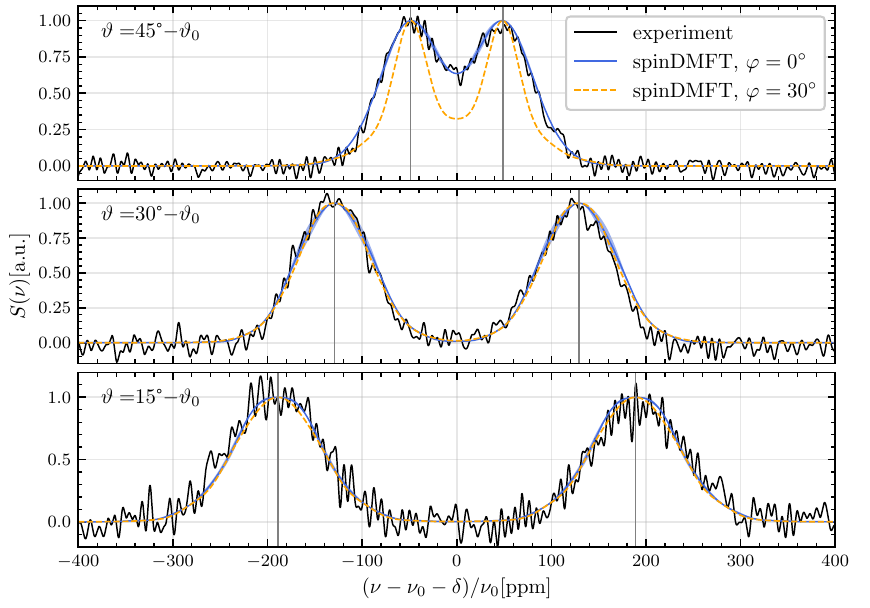}
  \caption{FID spectra of \ce{^{14}N} spins in an AlN monocrystal for different crystal rotation angles $\vartheta$. The solid black lines correspond to the experimental data from \refcite{zeman20}. The experimentally determined correction angle is given by $\vartheta_0 = \SI{-0.74}{\degree}$ and the Larmor frequency of \ce{^{14}N} is $\nu_0=\SI{28.905}{\mega\hertz}$. The blue solid lines correpond to the simulation results obtained by spinDMFT for $\varphi=\SI{0}{\degree}$, i.e., a rotation axis perpendicular to c and a or b. The numerical errors of spinDMFT are indicated by the transparent blue $1\sigma$-interval. By the vertical grey lines, we indicate the expected quadrupolar transition frequencies (without second-order secular correction). Note that the chemical shift has been ignored in the simulations and therefore needs to be removed from the experimental data for a proper comparison. We consider $\delta_{\text{iso}}=\SI{-292.6}{ppm}$ and $\delta_{\Delta}=\SI{-1.9}{ppm}$ \cite{zeman20} and subtract $\delta = \delta_{\text{iso}} - \delta_{\Delta} (3\cos^2(\vartheta)-1)/2$ from the experimental frequencies. For demonstration purposes, we also include the spinDMFT result for $\vartheta=\SI{45}{\degree}, \varphi=\SI{30}{\degree}$ as a dashed line. This underlines our claim that the azimuthal angle $\varphi$ has a visible influence on the spectra.}
  \label{fig:N_phi=30}
\end{figure}

The results for nitrogen are shown in \cref{fig:N_phi=30} for different crystal rotation angles $\vartheta$. The experimental data are shown by black and the spinDMFT data by blue solid lines. The intensity of both spectra are normalized with respect to the peak maxima. The agreement between theory and experiment is excellent for all considered rotation angles $\vartheta$. The peak positions are located at the expected quadrupolar transition frequencies, which we indicate by vertical grey lines. While this matching is not surprising, the accurate prediction of the line shapes by spinDMFT is indeed remarkable. We emphasize that the lineshapes are still close to Gaussian functions. Deviations from this behavior partially result from the superposition of the results from two distinct nitrogen sites. To highlight the relevance of the second orientation angle $\varphi$, we also present the result for $\varphi=\SI{30}{\degree}$ as dashed lines. Especially for $\vartheta=\SI{45}{\degree}$, the line shapes strongly differ from the ones for the $\varphi=\SI{0}{\degree}$ case.

\begin{figure}[h!]
  \centering
  \includegraphics[width=\textwidth]{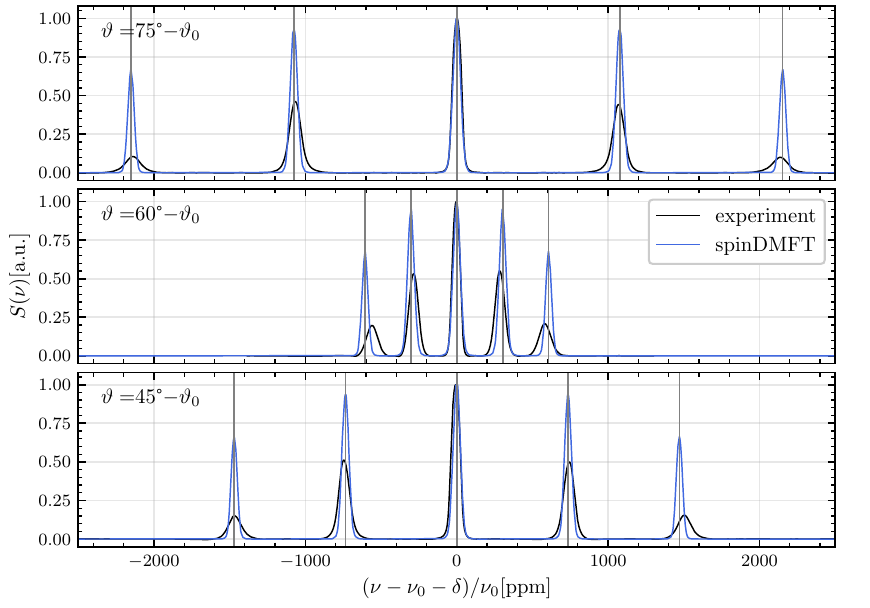}
  \caption{FID spectra of \ce{^{27}Al} spins in an AlN monocrystal for different crystal rotation angles $\vartheta$. The solid black lines correspond to the experimental data from \refcite{zeman20}. The experimentally determined correction angle is given by $\vartheta_0 = \SI{0.65}{\degree}$ and the Larmor frequency of \ce{^{27}Al} is $\nu_0=\SI{104.263}{\mega\hertz}$. The blue lines correpond to the simulation results obtained by spinDMFT. The numerical errors of spinDMFT are smaller than the widths of the lines. By the vertical grey lines, we indicate the expected quadrupolar transition frequencies (without second-order secular correction). Note that the chemical shift has been ignored in the simulations and therefore needs to be removed from the experimental data for a proper comparison. We consider $\delta_{\text{iso}}=\SI{113.6}{ppm}$ and $\delta_{\Delta}=\SI{12.7}{ppm}$ \cite{zeman20} and subtract $\delta = \delta_{\text{iso}} - \delta_{\Delta} (3\cos^2(\vartheta)-1)/2$ from the experimental frequencies.}
  \label{fig:Al_phi=30}
\end{figure}

In \cref{fig:Al_phi=30}, we present the results for the aluminium spectra. Here, the situation is different. First of all, we note that the variation of the azimuthal angle $\varphi$ has a non-visible effect on the line shapes. This is because the gyromagnetic ratio of the aluminium spins is larger by a factor of almost 4 with respect to the nitrogen spins. Thus, the aluminium lineshapes are dominanted by the much stronger Al-Al couplings, which do not depend on $\varphi$, as can be seen, for example, in the matrices in \cref{tab:quad_coup_matrix}. The agreement between theory and experiment is very good for the central peak. However, the satellite transition peaks differ in height, width and even slightly in position. The small discrepancy in the positions can be attributed to second-order effects from the secular approximation, since the quadrupolar coupling of aluminium is fairly large. This is thoroughly discussed in \refcite{zeman20}. The origin of the differences in height and width is less clear. We list a few possible reasons in the following.

First, we mention that the simulated spectra correspond only to the spin autocorrelations and not the FID. As discussed before, the FID contains also pair correlations, which typically broaden the line shapes. However, while such an effect becomes significant for FID's of homonuclear $S=\sfrac12$ ensembles \cite{graes24}, we speculate that its influence is less pronounced under strong quadrupolar interactions because the latter render the spin dynamics more local. Moreover, we would expect such a deviation to be visible also in the central peak, which is not the case here. Second, the mean-field approximation itself can lead to deviations. However, the aluminium spectrum is dominated by Al-Al couplings and each aluminium spin has already 12 nearest neighbors of the same species. This should perfectly justify spinDMFT, which was shown to work excellently already for less dense and less local spin systems \cite{graes21,graes24}. 

For these reasons, we consider the modelling rather than the mean-field approximation to be the source of the discrepancies. One possible reason could be a variation of the quadrupolar coupling over the lattice due to impurities or mosaicity. Slight variations of $\chi_{\Al}$ would broaden the satellite peaks and explain the observed deviation partially. Another factor could be that the nutation frequencies of the satellite peaks deviate from those of the central peak. The employed pulses are typically optimized to the central peak leading to reduced signal intensities of the satellite transitions. Finally, we also mention that second order effects from the secular approximation affect not only the peak position, but also the line shapes. Second-order corrections may be calculated and included in the simulations in future studies.

\section{Comparison to classical dynamics}
\label{sec:classspinDMFT}

As pointed out in \refcite{graes21}, the dynamics of a single spin in a classical mean-field is essentially classical. This is because the spin's equation of motion is linear in spin operators which makes it equivalent to that of a classical spin according to Ehrenfest's theorem. However, with the bilinear quadrupolar term included, this conclusion does not hold anymore. Despite being completely local, spinDMFT captures beyond-classical behavior for finite quadrupolar interactions.

To highlight the importance of simulating the local degrees of freedom quantum mechanically, we perform a simulation of the classical analogue system for comparison. The classical equation of motion can be derived as in \refcite{david15} yielding
\begin{align}
    \frac{\partial\vec{S}}{\partial t} &= \frac{\partial H}{\partial \vec{S}} \times \vec{S}.
\end{align}
Inserting the mean-field Hamiltonian from \cref{eqn:mfHam}, we obtain 
\begin{align}
    \frac{\partial\vec{S}}{\partial t} &= \vec{V}(t) \times \vec{S} + 6 \Omega S^{z} 
    \begin{pmatrix}
    -S^y \\
    S^x \\
    0 \\
    \end{pmatrix},
\end{align}
which is equivalent to the quantum equation of motion, when replacing $\vec{S} \to \vS$ and symmetrizing the last term. To simulate the classical dynamics, we average over the mean-field at all times as well as over the initial spin values at $t=0$. The mean-field average works in the same way as for the quantum case. For the spin average, we fix the spins length to $\sqrt{S(S+1)}$ ensuring 
\begin{align}
    \overline{S^{\alpha}\dsq} &= \langle \Sa\dsq \rangle
\end{align}
and choose the initial orientation uniformly distributed over the Bloch sphere due to the high-temperature limit.

\begin{figure}[h]
    \centering
    \includegraphics[width=\textwidth]{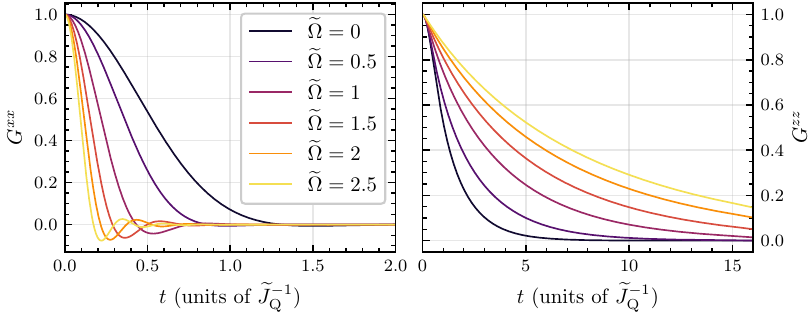}
    \caption{Universal results of the spin autocorrelations from classical spinDMFT presented in the main text. The spin length is incorporated in the time axis which is given in inverse units of the rescaled quadratic coupling constant $\widetilde{J}_{\mathrm{Q}}$. Numerical errors are of the order $\SI{1}{\percent}$ or smaller of the signal amplitude at $t=0$.}
    \label{fig:spinDMFTclass}
\end{figure}

The difference between quantum and classical dynamics already becomes apparent when considering the effect of a varied spin length $S$. The obtained equation of motion can be rewritten as
\begin{align}
    \frac{\partial\vec{n}}{\partial \widetilde{t}} &= \vec{W}(\widetilde{t}) \times \vec{n} + 6 \Omega n^{z} 
    \begin{pmatrix}
    -n^y \\
    n^x \\
    0 \\
    \end{pmatrix}
    \label{eqn:classEoMrenormalized}
\end{align}
using the renormalized quantities
\begin{align}
    \vec{n} &\coloneqq \frac{1}{\sqrt{S(S+1)}} \vec{S}, &
    \vec{W} &\coloneqq \frac{1}{\sqrt{S(S+1)}} \vec{V}, &
    \widetilde{t} &\coloneqq \sqrt{S(S+1)} t.
\end{align}
Note that a renormalization of $\vec{S}$ automatically implies a renormalization of $\vec{V}$ due to the self-consistency condition. The consequence of \cref{eqn:classEoMrenormalized} is that the dynamics is independent of the spin length except for a rescaling of the time axis. This fact is already in stark contrast to the quantum results shown in the previous section, where changing the spin length leads to qualitatively different behavior beyond a sheer scaling factor. The universal classical results for arbitrary spin length are presented in \cref{fig:spinDMFTclass} versus the time. Both the transverse and longitudinal autocorrelations clearly differ from the quantum results shown in \cref{fig:spinDMFTtimeresults}. The difference to the quantum mechanical results is the smallest for the largest value of the spin, $S=2$. This is not surprising because larger spins tend to behave more classically.

\newpage

\begin{figure}[h]
    \centering
    \includegraphics[width=\textwidth]{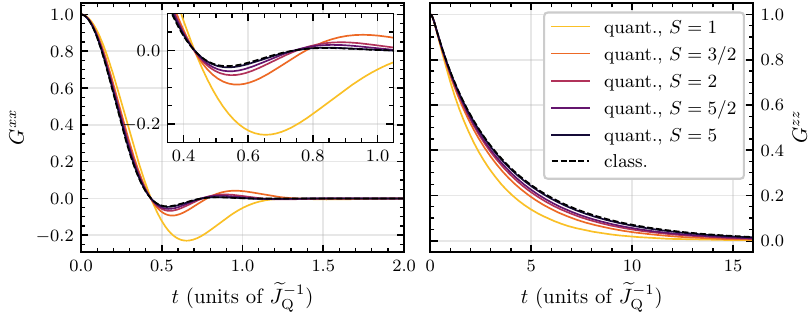}
    \caption{Comparison of quantum spinDMFT for different spin lengths with the universal result of classical spinDMFT for a fixed quadrupolar coupling strength $\widetilde{\Omega}=1$. Numerical errors are of the order $\SI{0.5}{\percent}$ or smaller of the signal amplitude at $t=0$.}
    \label{fig:large_spin_limit}
\end{figure}

This can be underlined analytically by the Frobenius norm
\begin{align}
    \norm{\op{A}} \coloneqq \frac1{d} \Tr{\op{A}^{\dagger} \op{A}},
\end{align}
of an operator $\op{A}$, where $d$ denotes the Hilbert space dimension. Similar to the consideration in \refcite{stane13}, we compare the norm of the commutator of two spin components
\begin{align}
    \norm{\left[\Sa,\Sb\right]} &= \norm{\Sa} = \frac{S(S+1)}{3}, & \alpha&\neq \beta,
\end{align}
to the norm of a product of two spin components
\begin{align}
    \norm{\Sa\Sb} &= \frac1{15} S(S+1) \left(S(S+1) + \frac12\right), & \alpha&\neq \beta.
\end{align}
We find that the relative commutator norm is suppressed by
\begin{align}
    \frac{\norm{\left[\Sa,\Sb\right]}}{\norm{\Sa\Sb}} &\propto \frac1{S^2}.
\end{align}
Hence, the error from ignoring the non-commutativity of spin operators becomes smaller and smaller if $S$ is increased resembling more and more classical behavior. As can be seen in \cref{fig:large_spin_limit}, this behavior is confirmed numerically by a direct comparison of the quantum results for increasing spin length with the classical ones.

\begin{figure}[h]
    \centering
    \includegraphics[width=\textwidth]{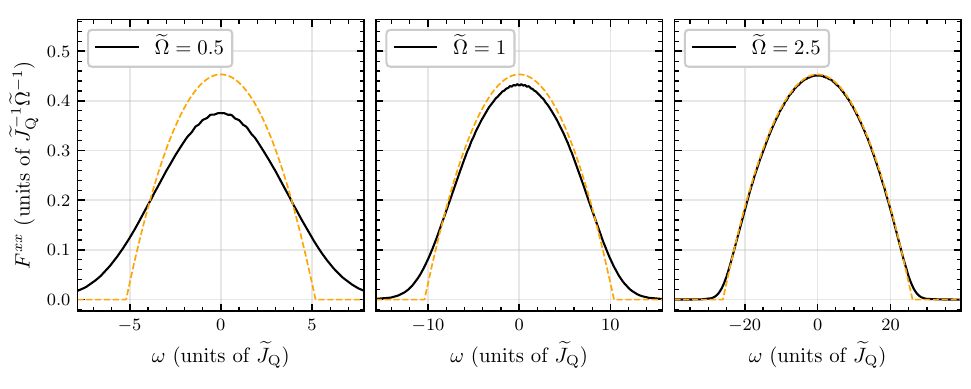}
    \caption{Fourier transform $F^{xx}(\omega)\coloneqq \int_{-\infty}^{\infty} \mathrm{e}^{-\im \omega t} G^{xx}(t) \mathrm{d}t$ of the classical transverse spin autocorrelation $G^{xx}(t)$ for different quadrupolar interaction strengths. The spin length is incorporated in the units through $\widetilde{J}_{\text{Q}}$. The signal $F^{xx}$ is shown in units of $\widetilde{J}_{\text{Q}}^{-1}\widetilde{\Omega}^{-1}$ to enhance the comparability of the results. The orange dashed lines indicate the exact classical result for large $\Omega$ according to \cref{eqn:classical_spectrum_large_Omega}. The numerical errors of the simulation data are smaller than the width of the lines.}
    \label{fig:classical_spectra}
\end{figure}

In \cref{fig:classical_spectra}, we provide the classical spectra of the transverse autocorrelation. As the classical spin is continuous, there is an infinite number of possible transitions so that even for large $\Omega$, no peak structure is obtained in contrast to the quantum case. Instead, one obtains a superposition of an infinite number of peaks. For $\Omega\ll \sigma$, these peaks are very close to one another which leads to shapes strongly ressembling Gaussian curves in total, since the individual peaks are Gaussian. For $\Omega\gg \sigma$, the shape of an individual peak becomes unimportant and only its position and weight matters. 

\FloatBarrier

Starting from the exact local result for quantum spins (see \cref{eqn:fitfunction} with the Gaussian functions replaced by $\delta$-distributions), it can be shown that
\begin{subequations}
\begin{align}
    F^{xx}_{\text{class}}(\omega) &= \frac{3}{S(S+1)} \pi \sum_{i=1}^{2S} \frac{i}{2} \left(1-\frac{i}{2S+1} \right) \nonumber \\ 
    &\qquad\qquad\qquad \times \Bigl[ \delta\bigl(\omega-3\Omega(2S-2i+1)\bigr) + \delta\bigl(\omega+3\Omega(2S-2i+1)\bigr) \Bigr] \label{eqn:classspectrum_a}\\
    &= \frac{\pi (2S+1)}{8\Omega S(S+1)} \sum_{\substack{x=1-2S\\ \text{odd}}}^{2S-1} \left[1-\left(\frac{x}{2S+1}\right)^2 \right] \left[ \delta\Bigl(\frac{\omega}{3\Omega}-x\Bigr) + \delta\Bigl(\frac{\omega}{3\Omega}+x\Bigr) \right] \label{eqn:classspectrum_b}\\
    &= \frac{\pi (2S+1)}{16\Omega S(S+1)} \int_{1-2S}^{2S-1} \mathrm{d}x \left[1-\left(\frac{x}{2S+1}\right)^2 \right]
    \left[ \delta\left(\frac{\omega}{3\Omega}-x\right) + \delta\left(\frac{\omega}{3\Omega}+x\right) \right] \label{eqn:classspectrum_c} \\
    &= \frac1{\widetilde{J}_{\text{Q}}\widetilde{\Omega}} \frac{\pi}{4\sqrt{3}} \left[ 1-\left( \frac{\omega/\widetilde{J}_{\text{Q}}}{6\sqrt{3} \widetilde{\Omega}} \right)^2 \right] \vartheta\left( 1 - \frac{|\omega|/\widetilde{J}_{\text{Q}}}{6\sqrt{3} \widetilde{\Omega}} \right).
\end{align}
\label{eqn:classical_spectrum_large_Omega}%
\end{subequations}
In the first step, we altered the sum index according to $x=2i-2S-1$; note that $x$ is altered in steps of 2. In the second step, we approximated the sum by an integral considering the limit of large $S$. Finally, we evaluated the integral and omitted any terms that are subdominant in $S$. This analytical result is shown in \cref{fig:classical_spectra} by the orange dashed line. In case of $\Omega\gg \sigma$, the classical spectrum consists of a single parabola instead of distinct Gaussian peaks. This further underlines the need for a quantum-mechanical simulation of quadrupolar systems. A calculation treating all spins as classical vectors can capture the spin dynamics only for large spins at best.


\section{Conclusion}
\label{sec:conclusion}

Spin dynamic mean-field theory (spinDMFT) is an efficient numerical approach to compute the dynamics of completely disordered dipolar spin systems. The key idea is to replace the environment of a spin by a time-dependent mean-field which is Gaussian distributed. This allows one to define a single-site problem and a self-consistency condition, which connects the variances of the mean-field to the spin autocorrelations. In this article, we showed how a quadrupolar interaction can be incorporated in spinDMFT: Since the quadrupolar term is completely local, it can be directly and exactly added to the single-site model. 

The numerical evaluation yields an exponential longitudinal relaxation and an oscillating transverse relaxation. The latter is best understood in the frequency spectrum, where peaks can be identified at the expected quadrupolar transitions with $\Delta m = \pm 1$. According to spinDMFT, the dipolar interaction broadens the resonance lines to Gaussian functions over the full range of parameters, i.e., spin lengths and quadrupolar interaction strengths. Remarkably, a fit with a single parameter, namely, the peak standard deviation, suffices for an adequate description of the spectrum. 

Real systems are usually more complex and involve for example different spin species. We demonstrated the versatility of spinDMFT by simulating the spin dynamics of quadrupolar \ce{^{14}N} and \ce{^{27}Al} nuclei in an AlN monocrystal. The obtained spectra were compared to experimental data for various crystal orientations. For \ce{^{14}N}, the agreement between theory and experiment is excellent for all orientations, which strongly supports spinDMFT as a prediction tool for dipolar broadening in quadrupolar spectra. Discrepancies are only visible in the \ce{^{27}Al} satellite peaks and can be attributed to model imperfections such as variations of the quadrupolar coupling over the lattice. In future works, such effects may be included in the approach.

Another goal of this article was to draw a comparison to the classical analogue system. While classical simulations often capture purely dipolar systems well, it turns out that the presence of a quadrupolar interaction precludes a classical description. The local spin degrees of freedom need to be simulated quantum-mechanically.

Typically, dipolar interactions are considered deleterious in NMR experiments due to the induced line broadening and associated difficulties in obtaining information from the spectra. However, if the dipolar line broadening can be predicted, systems with moderate dipolar interactions become accessible. As we demonstrated in this article, spinDMFT can be a suitable prediction tool for this scenario. We emphasize that the computational effort of spinDMFT is small allowing for extensive parameter sweeps and/or various extensions to increase the accuracy of the approach and access more complex systems. The mean-field framework can be extended to include inhomogeneous systems as well as explicit time dependencies, such as pulses.

In future works, spinDMFT could be extended to magic-angle spinning (MAS) in order to study residual dipolar broadening \cite{malae19,chave21}. This would also allow comparison to and prediction of experiments measuring MAS spectra of quadrupolar nuclei \cite{meado82,eliav13}. Since quadrupolar interactions and explicit time-dependencies are accessible, spinDMFT could also be a useful simulation tool to study motion in \ce{Li}-ion conductors \cite{boehm07,store14}. The results presented pave the way to a quantitative analysis and understanding of NMR results in a large variety of experiments.

\paragraph{Acknowledgments/Funding information}

We are grateful to T. Bräuniger for the provision of the experimental data. Furthermore, we thank M. Ernst, T. Bräuniger, L. Niccoli and R. Böhmer for useful discussions. We acknowledge funding by the Deutsche Forschungsgemeinschaft (DFG) in project UH90/14–2.

\paragraph{Code and data availability}

A code collection concerning spinDMFT and its extensions has been published under \url{https://doi.org/10.17877/TUDODATA-2025-MD4EYWOL}. The simulation data related to this manuscript can be found under \url{https://doi.org/10.5281/zenodo.17871931}.

\begin{appendix}

\end{appendix}


\nolinenumbers

\end{document}

%% file: commands.tex
\newcommand{\eff}{\mathrm{eff}}         
\newcommand{\mf}{\mathrm{mf}}           
\newcommand{\Qu}{\mathrm{Q}}            
\newcommand{\Al}{\mathrm{Al}}            
\newcommand{\Ni}{\mathrm{N}}            
\newcommand{\X}{\mathrm{X}}             

\newcommand{\mb}{\mathbf}
\newcommand{\dagg}{\dagger}

\def\op#1{\mb{#1}}

\newcommand{\oH}{\op{H}}

\newcommand{\oU}{\op{U}}
\newcommand{\oUd}{\op{U}^{\dagg}}

\newcommand{\oS}{\op{S}}
\newcommand{\Sx}{\oS^{x}}
\newcommand{\Sy}{\oS^{y}}
\newcommand{\Sz}{\oS^{z}}

\newcommand{\Sa}{\oS^{\alpha}}
\newcommand{\Sb}{\oS^{\beta}}

\newcommand{\Iz}{\op{I}^{z}}
\newcommand{\Ia}{\op{I}^{\alpha}}

\newcommand{\Sxi}{\Sx_{i}}
\newcommand{\Syi}{\Sy_{i}}
\newcommand{\Szi}{\Sz_{i}}

\newcommand{\Sxj}{\Sx_{j}}
\newcommand{\Syj}{\Sy_{j}}
\newcommand{\Szj}{\Sz_{j}}

\newcommand{\vS}{\vec{\op{S}}}
\newcommand{\vSi}{\vec{\op{S}}_{i}}
\newcommand{\vSj}{\vec{\op{S}}_{j}}
\newcommand{\vI}{\vec{\op{I}}}

\newcommand{\oV}{\op{V}}

\newcommand{\vV}{\vec{\oV}}

\newcommand{\cVz}{V^{z}}
\newcommand{\cVa}{V^{\alpha}}
\newcommand{\cVb}{V^{\beta}}

\newcommand{\cvV}{\vec{V}}

\def\dul#1{\underline{\underline{#1}}}

\def\expval#1{\langle#1\rangle}

\def\exp#1{\mathrm{e}^{#1}}

\newcommand{\JQ}{J_{\Qu}}

\newcommand{\zeff}{z_{\eff}}

\newcommand{\im}{\texttt{i}}

\newcommand{\nB}{\vec{n}_{B}}

\def\mfav#1{\overline{#1}^{\mf}}

\def\blue#1{{\color{blue}{#1}}}

\xdefinecolor{tugreen}{RGB}{132, 184, 25}
\xdefinecolor{tuorange}{RGB}{255, 175, 0}
\xdefinecolor{tublue}{RGB}{65, 105, 225}

\newlength{\textarrow}
\usepackage{tikz} 

\usepackage{tikz} 

%% file: transitions.pdf_tex
\begingroup%
  \makeatletter%
  \providecommand\color[2][]{%
    \errmessage{(Inkscape) Color is used for the text in Inkscape, but the package 'color.sty' is not loaded}%
    \renewcommand\color[2][]{}%
  }%
  \providecommand\transparent[1]{%
    \errmessage{(Inkscape) Transparency is used (non-zero) for the text in Inkscape, but the package 'transparent.sty' is not loaded}%
    \renewcommand\transparent[1]{}%
  }%
  \providecommand\rotatebox[2]{#2}%
  \newcommand*\fsize{\dimexpr\f@size pt\relax}%
  \newcommand*\lineheight[1]{\fontsize{\fsize}{#1\fsize}\selectfont}%
  \ifx\svgwidth\undefined%
    \setlength{\unitlength}{420.39821936bp}%
    \ifx\svgscale\undefined%
      \relax%
    \else%
      \setlength{\unitlength}{\unitlength * \real{\svgscale}}%
    \fi%
  \else%
    \setlength{\unitlength}{\svgwidth}%
  \fi%
  \global\let\svgwidth\undefined%
  \global\let\svgscale\undefined%
  \makeatother%
  \begin{picture}(1,0.27324095)%
    \lineheight{1}%
    \setlength\tabcolsep{0pt}%
    \put(0,0){\includegraphics[width=\unitlength,page=1]{transitions.pdf}}%
    \put(0.24882626,0.16948216){\color[rgb]{0,0,0}\makebox(0,0)[lt]{\lineheight{1.25}\smash{\begin{tabular}[t]{l}$m=\pm 1$\end{tabular}}}}%
    \put(0.24882626,0.04460056){\color[rgb]{0,0,0}\makebox(0,0)[lt]{\lineheight{1.25}\smash{\begin{tabular}[t]{l}$m=0$\end{tabular}}}}%
    \put(0.07220805,0.25333119){\color[rgb]{0,0,0}\makebox(0,0)[lt]{\lineheight{1.25}\smash{\begin{tabular}[t]{l}$E_\text{Q}$\end{tabular}}}}%
    \put(0,0){\includegraphics[width=\unitlength,page=2]{transitions.pdf}}%
    \put(0.10432044,0.16948216){\color[rgb]{0,0,0}\makebox(0,0)[rt]{\lineheight{1.25}\smash{\begin{tabular}[t]{r}$3\Omega$\end{tabular}}}}%
    \put(0.10432044,0.04460056){\color[rgb]{0,0,0}\makebox(0,0)[rt]{\lineheight{1.25}\smash{\begin{tabular}[t]{r}0\end{tabular}}}}%
    \put(0.20600972,0.10704137){\color[rgb]{0,0,1}\makebox(0,0)[lt]{\lineheight{1.25}\smash{\begin{tabular}[t]{l}$3\Omega$\end{tabular}}}}%
    \put(0.1997681,0.25868329){\color[rgb]{0,0,0}\makebox(0,0)[t]{\lineheight{1.25}\smash{\begin{tabular}[t]{c}$S=1$\end{tabular}}}}%
    \put(0,0){\includegraphics[width=\unitlength,page=3]{transitions.pdf}}%
    \put(0.57351861,0.16948216){\color[rgb]{0,0,0}\makebox(0,0)[lt]{\lineheight{1.25}\smash{\begin{tabular}[t]{l}$m=\pm \frac32$\end{tabular}}}}%
    \put(0.57351861,0.04460056){\color[rgb]{0,0,0}\makebox(0,0)[lt]{\lineheight{1.25}\smash{\begin{tabular}[t]{l}$m=\pm \frac12$\end{tabular}}}}%
    \put(0.39690042,0.25333119){\color[rgb]{0,0,0}\makebox(0,0)[lt]{\lineheight{1.25}\smash{\begin{tabular}[t]{l}$E_\text{Q}$\end{tabular}}}}%
    \put(0,0){\includegraphics[width=\unitlength,page=4]{transitions.pdf}}%
    \put(0.439717,0.16948216){\color[rgb]{0,0,0}\makebox(0,0)[rt]{\lineheight{1.25}\smash{\begin{tabular}[t]{r}$\frac{27\Omega}{4}$\end{tabular}}}}%
    \put(0.42901286,0.04460056){\color[rgb]{0,0,0}\makebox(0,0)[rt]{\lineheight{1.25}\smash{\begin{tabular}[t]{r}$\frac{3\Omega}{4}$\end{tabular}}}}%
    \put(0.53070218,0.10704137){\color[rgb]{0,0,1}\makebox(0,0)[lt]{\lineheight{1.25}\smash{\begin{tabular}[t]{l}$6\Omega$\end{tabular}}}}%
    \put(0.52446052,0.25868329){\color[rgb]{0,0,0}\makebox(0,0)[t]{\lineheight{1.25}\smash{\begin{tabular}[t]{c}$S=\frac32$\end{tabular}}}}%
    \put(0,0){\includegraphics[width=\unitlength,page=5]{transitions.pdf}}%
    \put(0.8982105,0.2087307){\color[rgb]{0,0,0}\makebox(0,0)[lt]{\lineheight{1.25}\smash{\begin{tabular}[t]{l}$m=\pm 2$\end{tabular}}}}%
    \put(0.8982105,0.00892007){\color[rgb]{0,0,0}\makebox(0,0)[lt]{\lineheight{1.25}\smash{\begin{tabular}[t]{l}$m=0$\end{tabular}}}}%
    \put(0.72159267,0.25333119){\color[rgb]{0,0,0}\makebox(0,0)[lt]{\lineheight{1.25}\smash{\begin{tabular}[t]{l}$E_\text{Q}$\end{tabular}}}}%
    \put(0,0){\includegraphics[width=\unitlength,page=6]{transitions.pdf}}%
    \put(0.75370512,0.2087307){\color[rgb]{0,0,0}\makebox(0,0)[rt]{\lineheight{1.25}\smash{\begin{tabular}[t]{r}$12\Omega$\end{tabular}}}}%
    \put(0.75370511,0.00892007){\color[rgb]{0,0,0}\makebox(0,0)[rt]{\lineheight{1.25}\smash{\begin{tabular}[t]{r}$0$\end{tabular}}}}%
    \put(0.85539438,0.13201771){\color[rgb]{0,0,1}\makebox(0,0)[lt]{\lineheight{1.25}\smash{\begin{tabular}[t]{l}$9\Omega$\end{tabular}}}}%
    \put(0.84915283,0.25868329){\color[rgb]{0,0,0}\makebox(0,0)[t]{\lineheight{1.25}\smash{\begin{tabular}[t]{c}$S=2$\end{tabular}}}}%
    \put(0,0){\includegraphics[width=\unitlength,page=7]{transitions.pdf}}%
    \put(0.8982105,0.05887276){\color[rgb]{0,0,0}\makebox(0,0)[lt]{\lineheight{1.25}\smash{\begin{tabular}[t]{l}$m=\pm 1$\end{tabular}}}}%
    \put(0,0){\includegraphics[width=\unitlength,page=8]{transitions.pdf}}%
    \put(0.75370511,0.05887276){\color[rgb]{0,0,0}\makebox(0,0)[rt]{\lineheight{1.25}\smash{\begin{tabular}[t]{r}$3\Omega$\end{tabular}}}}%
    \put(0,0){\includegraphics[width=\unitlength,page=9]{transitions.pdf}}%
    \put(0.85896243,0.03568039){\color[rgb]{0,0,1}\makebox(0,0)[lt]{\lineheight{1.25}\smash{\begin{tabular}[t]{l}$3\Omega$\end{tabular}}}}%
    \put(0.55652965,0.01784022){\color[rgb]{0,0,1}\makebox(0,0)[lt]{\lineheight{1.25}\smash{\begin{tabular}[t]{l}$0$\end{tabular}}}}%
  \end{picture}%
\endgroup%